\documentclass[prl,aps,epsf,amsfonts,floats,twocolumn,amssymb,amsmath,groupedaddress,showpacs,floatfix,nofootinbib]{revtex4}

\usepackage{graphicx}
\usepackage[]{latexsym}
\usepackage{bm}

\newcommand{\be}{\begin{equation}}\newcommand{\ee}{\end{equation}}
\newcommand{\bea}{\begin{eqnarray}}\newcommand{\eea}{\end{eqnarray}}
\newcommand{\brr}{\begin{array}}\newcommand{\err}{\end{array}}
\newcommand{\bit}{\begin{itemize}}\newcommand{\eit}{\end{itemize}}
\newcommand{\ben}{\begin{enumerate}}\newcommand{\een}{\end{enumerate}}

\newcommand{\ba}{\begin{array}}
\newcommand{\ea}{\end{array}}

\def\lan{\langle}
\def\lf{\left}

\def\non{\nonumber}\def\ran{\rangle}

\def\ri{\right}

\def\om{\omega}

\def\1{{_{1}}}\def\2{{_{2}}}

\def\noHe0{:\;\!\!\;\!\!:H_e(0):\;\!\!\;\!\!:}
\def\noHm0{:\;\!\!\;\!\!:H_\mu(0):\;\!\!\;\!\!:}

\def\lan{\langle}
\def\lf{\left}

\def\non{\nonumber}
\def\ran{\rangle}

\def\ri{\right}

\def\om{\omega}

\def\1{{_{1}}}\def\2{{_{2}}}

\begin{document}

\title{Spontaneous Supersymmetry Breaking Induced by Vacuum Condensates}

\author{ Antonio Capolupo${}^{\natural}$\footnote{Corresponding author, e-mail address: capolupo@sa.infn.it}}
\author{Marco Di Mauro}
 \affiliation{${}^{\natural}$ Dipartimento di Ingegneria Industriale,
  Universit\'a di Salerno, Fisciano (SA) - 84084, Italy}

\pacs{03.70.+k, 11.10.-z, 11.30.Pb}

\begin{abstract}

We propose a novel mechanism of spontaneous supersymmetry breaking which relies upon an ubiquitous feature of Quantum Field Theory, vacuum condensates.
Such condensates play a crucial r\^{o}le in many phenomena.
Examples include Unruh effect, superconductors, particle mixing, and quantum dissipative systems.
We argue that in all these phenomena supersymmetry, when present, is spontaneously broken.
Evidence for our conjecture is given for the Wess--Zumino model, that can be considered an approximation to the supersymmetric extensions of the above mentioned systems. The magnitude of the effect is estimated for a recently proposed experimental setup based on an optical lattice.

\end{abstract}
\maketitle

Supersymmetry (SUSY) is a hypothetical symmetry of nature that has many nice aspects both from the purely theoretical and from the phenomenological point of view.
It relates any boson to a fermion (called superpartner) with the same mass and internal quantum numbers, and vice-versa.
Up to now, there is no direct evidence for the existence of SUSY at a fundamental level because no meaningful sign of the superpartners has ever been observed. Therefore, SUSY must be a broken symmetry, allowing for superparticles to be heavier than the corresponding Standard Model particles, otherwise it must be ruled out as a fundamental symmetry of nature. However, this does not prevent SUSY to be realized as an \emph{emergent} symmetry, e.g. in condensed matter systems, a possibility which has been recently object of much attention \cite{Yue Yang}.

Motivated by the fact that SUSY is not experimentally observed, intensive study has been devoted to the analysis of the possibility of SUSY breaking.
The first proposals exhibiting spontaneous SUSY breaking were given in \cite{Fayet:1974jb,O'Raifeartaigh:1975pr}, while its dynamical breaking has been first discussed in \cite{Witten:1981nf} (see \cite{Shadmi:1999jy} for a review).

 In an apparently separated research line, a number of specific physical problems, in which the existence of vacuum condensate plays a pivotal role,  have been studied. A partial list of such phenomena includes examples from QFT in
 various external fields (Unruh effect \cite{Unruh:1976db},  Schwinger effect \cite{Schwinger:1951nm}), condensed matter physics (BCS theory of superconductivity \cite{Bardeen:1957mv}, graphene physics \cite{Iorio:2010pv}, Thermo Field Dynamics \cite{Takahasi:1974zn}), particle physics and cosmology (neutrino \cite{Blasone:1995zc,Blasone:2002jv} and meson mixing \cite{Blasone:2001du,Capolupo:2004pt}, dark energy \cite{Capolupo:2006et,Capolupo:2008rz,Capolupo:2007hy}),  and the quantization of dissipative systems \cite{Celeghini:1991yv}.

Despite the diversity of these phenomena, all of them are characterized by the presence of vacuum condensates. Such condensates can be effectively described by using Bogoliubov transformations. The specific details of the mechanism that induces the condensates (arising either by an external field or by the underlying dynamics), are contained in the coefficients of such transformations, whose form is otherwise universal (sligthly more complicated is the case of particle mixing \cite{Capolupo:2010ek}, in which one has a Bogoliubov transformation nested in a rotation and the mixing angle represents a further parameter of the theory, but this complication has no effects on what we will say).

In the present paper we show that these most interesting issues, namely SUSY breaking and systems with condensed vacua, are intimately bound together. In fact, when the above listed phenomena are considered in a supersymmetric context, vacuum condensates appear to provide a new mechanism of spontaneous SUSY breaking.
As already mentioned, the Bogoliubov transformations formalism describes Thermo Field Dynamics (TFD) which is equivalent to the standard, thermal ensemble based, description of finite temperature field theory (see e.g. \cite{Das:1997gg}). Since it is well known that SUSY is spontaneously broken at any nonzero temperature \cite{Das:1997gg, Buchholz:1997mf}, the fact that the above listed phenomena are described by exactly the same formalism as TFD implies that SUSY is spontaneously broken in all these instances.

The setting we have in mind is the following.
We start by a situation in which SUSY is preserved at the classical (lagrangian) level and study the vacuum condensation effects which are common to all the above phenomena.
 To do that, we consider a Bogoliubov transformation acting
  simultaneously and with the same parameters on the bosonic and on the fermionic degrees of freedom in order not to break SUSY explicitly. Therefore, it makes sense to talk about its spontaneous breaking, and we conjecture that this phenomenon does in fact take place.
  What happens is that the presence of the condensates shifts the vacuum energy density to a nonvanishing value. Since, as well known (see e.g. the discussion in \cite{Witten:1981nf}), in any field theory which has manifest supersymmetry at the lagrangian level, a nonzero vacuum energy represents a sufficient condition for the spontaneous breaking of SUSY (as a simple consequence of the SUSY algebra), this implies that SUSY is spontaneously broken.

The above reasoning applies to all mentioned systems. Since some of them are of great phenomenological interest,
they could be considered as good candidates for displaying spontaneous SUSY (either fundamental or emergent) breaking.

A first example of situation in which such a process may happen has been studied in detail by the authors in Ref.\cite{Capolupo:2010ek} (see also \cite{Mavromatos:2010ni}) where it has been shown that a SUSY breakdown is induced by particle mixing.
\vspace{5mm}

Here we give some evidence for our conjecture by performing generic Bogoliubov transformations on a simple example of supersymmetric field theory, the free Wess--Zumino model, and showing that the resulting vacuum has nonvanishing energy density. As remarked above, this simple picture should give a good qualitative understanding of the vacuum of more complicated systems.

Before going on, we recall some basic facts about Bogoliubov transformations in QFT \cite{Umezawa:1993yq}. We treat the bosonic case; the fermionic case will be considered when the Wess--Zumino Lagrangian will be analyzed explicitly.

Consider a set of bosonic ladder operators $a_{\mathbf{k}}$. The canonical commutation relations (CCRs) are:
$
[a_{\mathbf{k}}, a^{\dagger}_{\mathbf{p}}]=\delta^{3}(\mathbf{k}-\mathbf{p})\,,
$
with all other commutators vanishing. The vacuum $|0\rangle$ is defined by $a_{\mathbf{k}}|0\rangle=0$, and a Fock space is built out of it in the well known way. Such a space is an irreducible representation of the algebra of the CCRs.

A Bogoliubov transformation has the form:
\bea\label{Atilde}
\tilde{a}_{\mathbf{k}}(\xi) = U_{\mathbf{k}} \, a_{\mathbf{k}} - V_{\mathbf{k}} \,  a^{\dagger}_{\mathbf{k}};
\eea
with the requirement that it be canonical, i.e. it must leave the CCRs invariant. This request translates in the condition $|U_{\mathbf{k}} |^2 - |V_{\mathbf{k}} |^2=1$, which implies that the coefficients have the general form $U_{\mathbf{k}}=e^{i\phi_{1\mathbf{k}}}\, \cosh \xi_{\mathbf{k}}$ and $V_{\mathbf{k}}=e^{i\phi_{2\mathbf{k}}}\, \sinh\xi_{\mathbf{k}}$, where $\xi_{\mathbf{k}}$ is the transformation parameter. The phase factors are irrelevant in what follows, so they can be discarded.

The transformation (\ref{Atilde}) can be rewritten in terms of a generator $J(\xi)$ as
$
\tilde{a}_{\mathbf{k}}(\xi) = J^{-1}(\xi)\,  a_{\mathbf{k}}\,  J(\xi)\, ,
$
where $J(\xi) $  is given by
$
J(\xi) = \exp \lf[\frac{i}{2}\sum_{\mathbf{k}} \xi_{\mathbf{k}}\lf(a_{\mathbf{k}}^2 - (a_{\mathbf{k}}^{\dagger})^2\ri)\ri]\,,
$
and has the property $J^{-1}(\xi)=J(-\xi)$. The vacuum $|\tilde{0}(\xi)\rangle$ relative to the transformed operator $\tilde{a}_{\mathbf{k}}(\xi)$ is defined by $\tilde{a}_{\mathbf{k}}(\xi)|\tilde{0}(\xi)\rangle=0$ and is related to the vacuum $|0\rangle$ by
$
|\tilde{0}(\xi)\rangle = J^{-1} (\xi)|0\rangle\,.
$
This is a unitary operation if $\mathbf{k}$ assumes a \textit{discrete} range of values, which happens if there is a finite or countable number of CCRs. If this is the case, the Fock spaces built on the two vacua are equivalent, that is, any vector in one space can be expressed in terms of a well defined sum of vectors in the other space. This is essentially the result of the well known Stone--von Neumann theorem of Quantum Mechanics. But if $\mathbf{k}$ assumes a continuous infinity of values, which is what happens in QFT, we find
\bea\non
|\tilde{0}(\xi) \rangle &=& \exp\lf[ -\delta(\mathbf{0})\int\, d^3 \mathbf{k}\; \log\cosh\xi_{\mathbf{k}} \ri]
 \\ &\times & \exp\lf[ \int\, d^3 \mathbf{k}\; \tanh\xi_{\mathbf{k}} (a_{\mathbf{k}}^{\dagger})^2\ri] |0\rangle \,,
\eea
which is not a unitary transformation any more. This shows explicitly that the vacuum $|\tilde{0}(\xi) \rangle$ cannot be expressed as a superposition of vectors in the Fock space built over $|0\rangle$. The same is true for the whole Fock space built over $|\tilde{0}(\xi)\rangle$, i.e. the two Fock spaces are unitarily inequivalent. Each state of the family $|\tilde{0}(\xi) \rangle$ therefore represents in principle a viable vacuum state for the theory.

Let us now analyze the effects of a Bogoliubov transformation in the Wess--Zumino model \cite{Wess:1973kz}. This field theory is described by the Lagrangian:
\bea \label{WS}\non
\mathcal{L} &=& \frac{i}{2} \bar{\psi}\gamma_{\mu}\partial^{\mu}\psi + \frac{1}{2}\partial_{\mu}S\partial^{\mu}S + \frac{1}{2}\partial_{\mu}P\partial^{\mu}P
\\ &-& \frac{m}{2}  \bar{\psi}\psi - \frac{m^2}{2} (S^2 + P^2),
\eea
where $\psi$ is a Majorana spinor field, $S$ is a scalar field and $P$ is a pseudoscalar field. As well known, this Lagrangian is invariant under supersymmetry transformations \cite{Wess:1973kz}.

The fields are quantized by expanding them in modes:
\bea
\psi(x)&=& \sum_{r=1}^2\int \frac {d^3\mathbf{k}}{(2\pi)^{\frac{3}{2}}}\,\, e^{i \mathbf{k}\mathbf{x}}\lf[u^r_{\mathbf{k}}\alpha^r_{\mathbf{k}}(t)
+ v^r_{-\mathbf{k}}\alpha^{r \dagger}_{\mathbf{-k}}(t)\ri],\\
S(x)&=& \int \frac {d^3\mathbf{k}}{(2\pi)^{\frac{3}{2}}}\,\, \frac{1}{\sqrt{2 \omega_{k}}} \,\, e^{i \mathbf{k}\mathbf{x}}\lf[b_{\mathbf{k}}(t)
+ b^{\dagger}_{\mathbf{-k}}(t)\ri],\\
P(x)&=& \int \frac {d^3\mathbf{k}}{(2\pi)^{\frac{3}{2}}}\,\, \frac{1}{\sqrt{2 \omega_{k}}} \,\, e^{i \mathbf{k}\mathbf{x}}\lf[c_{\mathbf{k}}(t)
+ c^{\dagger}_{\mathbf{-k}}(t)\ri],
\eea
where $v^r_{\mathbf{k}}=\gamma_0 C (u^r_{\mathbf{k}})^*$ and $u^r_{\mathbf{k}}=\gamma_0 C (v^r_{\mathbf{k}})^*$ since $\psi$ is a Majorana spinor, $\alpha^r_{\mathbf{k}}(t) = \alpha^r_{\mathbf{k}} e^{-i\omega_{k}t}$, $b_{\mathbf{k}}(t) = b_{\mathbf{k}} e^{-i\omega_{k}t}$, $c_{\mathbf{k}}(t) = c_{\mathbf{k}} e^{-i\omega_{k}t}$ and $\omega_{\mathbf{k}}=\sqrt{k^2 + m^2}$. The vacuum $|0\rangle=|0\rangle^{\psi}\otimes|0\rangle^S\otimes|0\rangle^P$ is defined as the state annihilated by the operators $\alpha^r_{\mathbf{k}}$, $b_{\mathbf{k}}$ and $c_{\mathbf{k}}$.

We consider the following three sets of Bogoliubov transformations acting on the ladder operators of the fermion and of the bosons:
\bea\label{Bog1}
\tilde{\alpha}^r_{\mathbf{k}}(\xi, t) &=& U^{\psi}_{\mathbf{k}} \, \alpha^r_{\mathbf{k}}(t) + V^{\psi}_{-\mathbf{k}} \, \alpha^{r\dagger}_{-\mathbf{k}}( t)\,, \non
\\
\tilde{\alpha}^{r\dagger}_{-\mathbf{k}}(\xi,  t) &=& U^{\psi *}_{-\mathbf{k}} \, \alpha^{r\dagger}_{-\mathbf{k}}(t) + V^{\psi *}_{ \mathbf{k}} \, \alpha^{r}_{\mathbf{k}}(t)\,,
\\[2mm]\non
\tilde{b}_{\mathbf{k}}(\eta,  t) &=& U^{S}_{\mathbf{k}} \, b_{\mathbf{k}}(t) - V^{S}_{-\mathbf{k}}  \,b^{\dagger}_{-\mathbf{k}}(t)\,,
\\\label{Bog2}
\tilde{b^{\dag}}_{-\mathbf{k}}(\eta,  t) &=& U^{S *}_{-\mathbf{k}} \, b^{\dag}_{-\mathbf{k}}(t) - V^{S *}_{\mathbf{k}}  \,b_{\mathbf{k}}(t)\,,
\\[2mm]\non
\tilde{c}_{\mathbf{k}}(\eta,  t) &=& U^{P}_{\mathbf{k}} \, c_{\mathbf{k}}(t) - V^{P}_{-\mathbf{k}}  \,c^{\dagger}_{-\mathbf{k}}(t)\,,
\\\label{Bog3}
\tilde{c^{\dag}}_{-\mathbf{k}}(\eta,  t) &=& U^{P *}_{-\mathbf{k}} \, c^{\dag}_{-\mathbf{k}}(t) - V^{P *}_{\mathbf{k}}  \,c_{\mathbf{k}}(t)\,,
\eea
in which the Bogoliubov coefficients of scalar and pseudoscalar bosons are equal each other,  $U^{S}_{\mathbf{k}} =U^{P}_{\mathbf{k}} $ and $V^{S}_{\mathbf{k}} =V^{P}_{\mathbf{k}} $. We denote such quantities as $U^{B}_{\mathbf{k}} $ and $V^{B}_{\mathbf{k}} $, respectively. The Bogoliubov coefficients of the fermions satisfy the constraints
$
U^{\psi}_{\mathbf{k}} = U^{\psi}_{-\mathbf{k}}$, $ V^{\psi}_{\mathbf{k}} = -V^{\psi}_{-\mathbf{k}}$, $|U^{\psi}_{\mathbf{k}}|^2 + |V^{\psi}_{\mathbf{k}}|^2 = 1\,,
$
while the ones relative to the bosons satisfy the conditions
$
U^{B}_{\mathbf{k}} = U^{B}_{-\mathbf{k}}$, $ V^{B}_{\mathbf{k}} = V^{B}_{-\mathbf{k}}$, $|U^{B}_{\mathbf{k}}|^2 - |V^{B}_{\mathbf{k}}|^2 = 1\,,
$
so they have the general form:
$
U^{\psi}_{\mathbf{k}} =e^{i\phi_{1\mathbf{k}}}\cos\xi_{\mathbf{k}}(\zeta)$, $ V^{\psi}_{\mathbf{k}} =e^{i\phi_{2\mathbf{k}}}\sin\xi_{\mathbf{k}}(\zeta)$, $
U^{B}_{\mathbf{k}} = e^{i\gamma_{1\mathbf{k}}}\cosh\eta_{\mathbf{k}}(\zeta)$, $ V^{B}_{\mathbf{k}} =e^{i\gamma_{2\mathbf{k}}}\sinh\eta_{\mathbf{k}}(\zeta)
$, respectively.
Here $\zeta$ represents the relevant parameter which controls the physics underlying the Bogoliubov transformation. For example, $\zeta$ is the temperature $T$ in   Thermo Field Dynamics case and $\zeta$ is the acceleration of the observer in Unruh effect case;
 moreover, as announced, since the phases $\phi_{i\mathbf{k}}$, $\gamma_{i\mathbf{k}}$, with $i=1,2$,  are irrelevant, we neglect them.

The  transformations (\ref{Bog1})--(\ref{Bog3}) can be written at any time $t$ in terms of the generator $J(\xi, \eta, t)$ as:
\bea
\tilde{\alpha}^r_{\mathbf{k}}( \xi,  t) &=& J^{-1} (\xi,\eta,  t)\,\alpha^r_{\mathbf{k}}(t) J(\xi,\eta,  t)\,,
\eea
and similar relations hold for the other annihilation and creation operators; the generator is
$
J(\xi,\eta,   t) = J_{\psi}(\xi,  t) J_{S}(\eta,  t) J_{P}(\eta,  t)\,,
$
where
\bea\non
J_{\psi} &=& \exp \lf[\frac{1}{2}\int d^{3} \mathbf{k} \, \xi_{\mathbf{k}}(\zeta)\lf(\alpha^r_{\mathbf{k}}(t) \alpha^r_{-\mathbf{k}}(t)  - \alpha^{r \dagger}_{-\mathbf{k}}(t) \alpha_{ \mathbf{k}}^{r \dagger}(t)  \ri)\ri],
\\\non
J_{S} &=& \exp \lf[-i\int d^{3} \mathbf{k}\, \eta_{\mathbf{k}}(\zeta)\lf(b_{\mathbf{k}}(t) b_{-\mathbf{k}}(t)  - b_{-\mathbf{k}}^{\dagger}(t) b_{ \mathbf{k}}^{\dagger}(t)  \ri)\ri],
\\\non
J_{P} &=& \exp \lf[-i\int d^{3} \mathbf{k}\, \eta_{\mathbf{k}}(\zeta)\lf(c_{\mathbf{k}}(t) c_{-\mathbf{k}}(t)  - c_{-\mathbf{k}}^{\dagger}(t) c_{ \mathbf{k}}^{\dagger}(t)  \ri)\ri].
\\
\eea

The vacuum annihilated by the new annihilators is
$|\tilde{0}(   t)\rangle=|\tilde{0}(   t)\rangle_{\psi}\otimes |\tilde{0}(  t)\rangle_{S}\otimes |\tilde{0}(   t)\rangle_{P}$, where the new vacua for fermion and bosons fields $|\tilde{0}(  t)\rangle_{\alpha}$, with $\alpha = \psi , S , P$, are related to the original ones $|0\rangle_{\alpha}$ by the relations
$
|\tilde{0}(  t)\rangle_{\psi} = J^{-1}_{\psi}(\xi,  t)|0\rangle_{\psi} $,
$ |\tilde{0}(  t)\rangle_{S} = J^{-1}_{S}(\eta,  t)|0\rangle_{S}$,
$  |\tilde{0}(   t)\rangle_{P} = J^{-1}_{P}(\eta,  t)|0\rangle_{P}
$, respectively,
and then
\bea
|\tilde{0}(   t)\rangle = J^{-1}(\xi,\eta,  t)|0\rangle\,.
\eea
The vacuum
$|\tilde{0}( t)\rangle$  is the physical vacuum for the systems listed above studied in a supersymmetric contest
(the parameter $\zeta$ and the form of the Bogoliubov coefficients specify the particular phenomenon). $|\tilde{0}(   t)\rangle$  is a condensate of couples of particles and antiparticles. This can be seen explicitly by looking at the condensation densities of fermions and bosons, which are given by
\bea\label{cond1}
 &&\langle\tilde{0}( t)| \alpha^{r\dagger}_{\mathbf{k}} \alpha^r_{\mathbf{k}} |\tilde{0}(  t)\rangle  = |V_{\mathbf{k}}^{\psi}|^2\,;
\\
&&\langle\tilde{0}(   t)| b^{\dagger}_{\mathbf{k}} b_{\mathbf{k}} |\tilde{0}(   t)\rangle  =  \langle\tilde{0}( t)| c^{\dagger}_{\mathbf{k}} c_{\mathbf{k}} |\tilde{0}( t)\rangle = |V_{\mathbf{k}}^{B}|^2;\,\,\,\,\,\,\, \label{cond2}
\eea
where Eqs.(\ref{Bog1})--(\ref{Bog3}) have been used. The results of Eqs.(\ref{cond1}) and (\ref{cond2})  lead to an energy density different from zero for this vacuum.
Indeed, consider the free Hamiltonian $H$ corresponding to the Lagrangian in Eq.(\ref{WS}), $H= H_{\psi} +  H_B $ where $H_B = H_S + H_P$.
The expectation values of the fermion  and boson Hamiltonians on $|\tilde{0}(  t)\rangle$  are given by
\bea\non\label{Hpsiasp}
&&\langle\tilde{0}(  t)| H_{\psi} |\tilde{0}(  t)\rangle =
 \\\non
 &&= \sum_r\int  d^3\mathbf{k} \, \omega_{\mathbf{k}} \, \langle\tilde{0}(   t)| \lf(\alpha^{r\dagger}_{\mathbf{k}} \alpha^r_{\mathbf{k}} -\frac{1}{2}\ri) |\tilde{0}(   t)\rangle
 \\
&& = - \int \; d^3\mathbf{k}\; \omega_{\mathbf{k}} \,(1- 2 |V^{\psi}_{\mathbf{k}}|^2)\,,
\eea
and
\bea\label{HBasp}
\langle\tilde{0}(   t)| H_B |\tilde{0}(  t)\rangle = \int\; d^3\mathbf{k}\; \omega_{\mathbf{k}} (1 + 2 |V^{B}_{\textbf{k}}|^2)\,,
\eea
respectively. Then combining Eqs.(\ref{Hpsiasp}) and (\ref{HBasp}), we have the  final result
\bea\label{Ht}
\langle\tilde{0}(  t)| H |\tilde{0}(  t)\rangle = 2 \int\; d^3\mathbf{k}\; \omega_{\mathbf{k}} (|V^{\psi}_{\textbf{k}}|^2 + |V^{B}_{\textbf{k}}|^2)\,,
\eea
which is different from zero and positive, as we wanted to show\footnote{
Notice that, on the contrary, the expectation value of $H$ on the vacuum $|0 \rangle$ is zero, $\langle 0|H|0\rangle = 0 $, since the negative fermion vacuum energy,  $\langle 0|H_{\psi}|0\rangle = - \int d^3\mathbf{k}\,  \omega^{\psi}_{\mathbf{k}}$, is compensated by the positive boson vacuum energy, $\langle 0|H_{B}|0\rangle =  \int d^3\mathbf{k}\, \omega^{B}_{\mathbf{k}}$, being $\omega^{\psi}_{\mathbf{k}} =  \omega^{B}_{\mathbf{k}} =  \omega_{\mathbf{k}}$.}.

 As clear from
the above, at the origin of this result there are the fermionic and bosonic
condensates which both lift  the vacuum energy by a positive amount, while the bosonic and fermionic shifts should have opposite sign (and equal amplitude) to make it be zero.

 The clarification of some important points is now in order. First, a free model like the one discussed above gives only a rough approximation and captures only a part of the physics, which nevertheless seems to be universal\footnote{it is tempting here to make a comparison with the mean field approximation, which is rough and also not controllable, but gives results which are universal, in the sense of being independent of the underlying dynamics. The parallel is very strong in the case of BCS theory, in which the mean field Hamiltonian is diagonalized by means of a Bogoliubov transformation.}. The presence of the state $|0\rangle$, which can be misleading, is in fact an artifact of the approximation. In realistic systems this vacuum is unstable due to the dynamics (as is well known for example to be the case in BCS theory) or due to the presence of a supercritical field (in the Schwinger effect case), or simply it is not accessible to the observer (Unruh effect). The phenomenon of condensation takes place and the true vacuum is therefore the transformed one $|\tilde{0}(  t)\rangle$. In the supersymmetric context, this fact together with the nonvanishing energy density of this vacuum implies the spontaneous breaking of SUSY.

 The scale at which the SUSY spontaneous breaking occurs, which is related to the mass shift between each particle and its superpartner, obviously depends on the particular mechanism which triggers the condensation, which as repeatedly emphasized is by itself very model dependent.

\vspace{5mm}

The main result of this letter, Eq.(\ref{Ht}), holds for disparate physical phenomena. As remarked above, the explicit form of the Bogoliubov coefficients
$V^{\psi}_{\textbf{k}}$ and $V^{B}_{\textbf{k}}$
specifies the particular system.
For example, in the case of the Unruh effect, the Bogoliubov coefficients that allow to express the Minkowski vacuum in terms of Rindler states are $V^{B}_{\textbf{k}} = \sqrt{\frac{1}{{e^{2 \pi \omega_{\textbf{k}}/a}}-1}}$ and $V^{\psi}_{\textbf{k}} = \sqrt{\frac{1}{{e^{2 \pi \omega_{\textbf{k}}/a}}+1}}$. Here $a$ is the acceleration of the observer. For completeness we show the relation between the two vacua in the case of a single scalar field \cite{Crispino2008}:
\bea
|0\rangle_M \sim \exp \lf( \frac{1}{2}\sum_{\textbf{k}} e^{-\pi \omega_{\textbf{k}}/a} a^{\dagger}_R a^{\dagger}_L\ri)|0\rangle_R ,
\eea
where $R$ and $L$ refer to modes supported in the right and left Rindler wedges respectively.

In subnuclear physics, an important instance of SUSY breaking phenomenon is provided by particle mixing \cite{Capolupo:2010ek}. In the case of mixing of two fields with masses $m_1, m_2$ the Bogoliubov coefficients are $
V^{\psi}_{{\bf k}}  =  \frac{ (\om_{k,1}+m_{1}) - (\om_{k,2}+m_{2})}{2
\sqrt{\om_{k,1}\om_{k,2}(\om_{k,1}+m_{1})(\om_{k,2}+m_{2})}}\, |{\bf k}| \, ,
$
and
$ V^B_{{\bf k}}  =   \frac{1}{2} \lf( \sqrt{\frac{\om_{k,1}}{\om_{k,2}}} -
\sqrt{\frac{\om_{k,2}}{\om_{k,1}}} \ri)\,.
$
For the relation between the flavor vacuum and the free field vacuum see refs \cite{Blasone:1995zc, Blasone:2001du}.
In this case Eq.(\ref{Ht}) is replaced by
$
\, {}_{f}\lan 0| H | 0 \ran_{f}\,=\,
2\,\sin^{2}\theta \, \int d^{3}{\bf k} \,
(\omega_{k,1} + \omega_{k,2}) \,(|V^{\psi}_{\bf k}|^{2} + \,|V^B_{\bf k}|^{2})
\,,
$
where $| 0 \ran_{f}$ is the physical vacuum, called the flavor vacuum, $\theta$ is the mixing angle and the $\sin^{2}\theta$ factor is due to the additional rotation involved.

In the already cited case of Thermo Field Dynamics, the  parameter  $\zeta$ is the temperature. The finite temperature physics is described by a thermal vacuum, and one just recovers the well known result that SUSY is spontaneously broken at any nonzero temperature \cite{Das:1997gg,Buchholz:1997mf}.

A possibility to test our conjecture with table top experiments could be offered by forthcoming laser cooling experiments. Indeed, recently an experiment was proposed in which the Wess-Zumino model in $2+1$ dimensions can emerge from a mixture of cold atoms-molecules trapped in two dimensional optical lattices \cite{Yue Yang}. In this case SUSY is preserved at zero temperature and is broken at $T\neq 0$.
Then, a signature of SUSY breaking in such a system can be probed experimentally, not only by the detection of a thermal Goldstone fermion -- the phonino,  as suggested in Ref.\cite{Yue Yang}, but also by the detection of the constant background noise due to the nonzero energy of the thermal vacuum, given by (\ref{Ht}), with $V^{B}_{\textbf{k}} = \sqrt{\frac{1}{{e^{\beta \omega_{\textbf{k}}}}-1}}$ and $V^{\psi}_{\textbf{k}} = \sqrt{\frac{1}{{e^{\beta \omega_{\textbf{k}}}}+1}}$, where $\beta$ is the inverse temperature (in units such that $k_B=1$).
Considering a two-dimensional optical lattice and a tuning of the parameters such that the effective mass of the emerging fields is zero, the vacuum energy density due to Eq.(\ref{Ht})  is given by
$<H>\,=\,14 \, \pi\,\zeta(3)\, T^{3} $.

Another very interesting issue to investigate in such optical systems would be the possibility of inducing a non thermal spontaneous breaking of SUSY by means of a \emph{quantum} phase transition.

Moreover, a system which can display SUSY breaking at cosmological scales is represented by curved spacetimes which classically allow for the existence of global SUSY \cite{Festuccia:2011ws}, such as anti--deSitter space (see e.g. \cite{deWit:1999ui}). Studying this situation in detail will be the purpose of future work.

In conclusion, we have conjectured that in a supersymmetric field theory spontaneous breaking of SUSY can be a result of the presence of vacuum condensates. We have given evidence for this conjecture in the simple case of the free Wess--Zumino model, but the occurrence of vacuum condensates is very general, so we expect our result to be generalizable to more complex and physically relevant situations, in which actual breaking of SUSY can occur.

The results here proposed could also have cosmological implications. Indeed, the non-zero vacuum energy induced by condensates which appear in different physical phenomena
could be related to the dark  energy component of the universe. Further analysis of these
aspects will be done in a forthcoming paper.

The authors would like to thank Dr. A. Naddeo  for useful discussions
and encouragements. Partial financial support
from MIUR is acknowledged.


\begin{thebibliography}{999}

\bibitem{Yue Yang}
Y.~Yu, K.~Yang
{\it    Phys. Rev. Lett.} {\bf 105}, 150605 (2010).


\bibitem{Fayet:1974jb}
  P.~Fayet, J.~Iliopoulos,
 {\it   Phys.\ Lett.\  B} {\bf 51}, 461 (1974).

\bibitem{O'Raifeartaigh:1975pr}
  L.~O'Raifeartaigh,
 {\it   Nucl.\ Phys.\  B} {\bf 96}, 331 (1975).

\bibitem{Witten:1981nf}
  E.~Witten,
 {\it   Nucl.\ Phys.\  B} {\bf 188}, 513 (1981).

\bibitem{Shadmi:1999jy}
  Y.~Shadmi, Y.~Shirman,
 {\it   Rev.\ Mod.\ Phys.\ } {\bf 72}, 25 (2000).





\bibitem{Unruh:1976db}
  W.~G.~Unruh,
{\it    Phys.\ Rev.\  D} {\bf 14}, 870 (1976).

\bibitem{Schwinger:1951nm}
  J.~S.~Schwinger,
 {\it   Phys.\ Rev.\ } {\bf 82}, 664 (1951).


\bibitem{Bardeen:1957mv}
  J.~Bardeen, L.~N.~Cooper, J.~R.~Schrieffer,
 {\it   Phys.\ Rev.\   } {\bf 108}, 1175 (1957).

\bibitem{Iorio:2010pv}
  A.~Iorio,
 {\it   Annals Phys.\ } {\bf 326}, 1334 (2011).

\bibitem{Takahasi:1974zn}
  Y.~Takahasi, H.~Umezawa,
{\it    Collect.\ Phenom.\ } {\bf 2}, 55 (1975).

\bibitem{Blasone:1995zc}
  M.~Blasone, P.~A.~Henning and G.~Vitiello,
 {\it      Phys.\ Lett.\ B} {\bf 451}, 140 (1999).


\bibitem{Blasone:2002jv}
  M.~Blasone, A.~Capolupo, G.~Vitiello,
 {\it   Phys.\ Rev.\ D } {\bf 66}, 025033 (2002).


\bibitem{Blasone:2001du}
  M.~Blasone, A.~Capolupo, O.~Romei, G.~Vitiello,
{\it    Phys.\ Rev.\ D} {\bf 63}, 125015 (2001).

\bibitem{Capolupo:2004pt}
  A.~Capolupo, C.~-R.~Ji, Y.~Mishchenko, G.~Vitiello,
 {\it   Phys.\ Lett.\ B} {\bf 594}, 135-140 (2004).


\bibitem{Capolupo:2006et}
  A. Capolupo,  S. Capozziello,  G. Vitiello,
 {\it  Phys.\ Lett.\ A} {\bf 363}, 53 (2007).


\bibitem{Capolupo:2008rz}
  A. Capolupo, S. Capozziello, G. Vitiello,
 {\it  Phys.\ Lett.\ A} {\bf 373}, 601 (2009).


\bibitem{Capolupo:2007hy}
  A.~Capolupo, S.~Capozziello, G.~Vitiello,
{\it   Int.\ J.\ Mod.\ Phys.\ A} {\bf 23}, 4979 (2008),
  %
  M.~Blasone, A.~Capolupo, S.~Capozziello, G.~Vitiello,
{\it   Nucl.\ Instrum.\ Meth.\  A} {\bf588}, 272 (2008),
%
  M.~Blasone, A.~Capolupo, G.~Vitiello,
{\it   Prog.\ Part.\ Nucl.\ Phys.\ } {\bf 64}, 451 (2010),
   M. Blasone,  A. Capolupo,  S. Capozziello,  S. Carloni,  G. Vitiello,
{\it  Phys.\ Lett.\ A}  {\bf 323}, 182 (2004).



\bibitem{Celeghini:1991yv}
  E.~Celeghini, M.~Rasetti, G.~Vitiello,
{\it    Annals Phys.\ } {\bf 215}, 156 (1992).





\bibitem{Capolupo:2010ek}
  A.~Capolupo, M.~Di Mauro, A.~Iorio,
 {\it   Phys.\ Lett.\  A} {\bf 375}, 3415 (2011).


\bibitem{Das:1997gg}
  A.~K.~Das, Finite temperature field theory,
World Scientific (Singapore) 1997.

\bibitem{Buchholz:1997mf}
  D.~Buchholz, I.~Ojima,
{\it    Nucl.\ Phys.\  B} {\bf 498}, 228 (1997).



\bibitem{Mavromatos:2010ni}
  N.~E.~Mavromatos, S.~Sarkar, W.~Tarantino,
 {\it   Phys.\ Rev.\  D} {\bf 84}, 044050 (2011).

\bibitem{Umezawa:1993yq}
  H.~Umezawa,
  Advanced field theory: Micro, macro, and thermal physics,
{\it  New York, USA: AIP} 238 p. (1993)

\bibitem{Wess:1973kz}
  J.~Wess, B.~Zumino,
 {\it   Phys.\ Lett.\  B} {\bf 49}, 52 (1974).


\bibitem{Crispino2008}
L.~C.~B.~Crispino, A.~ Higuchi, G.~E.~A.~ Matsas,
{\it Rev. Mod. Phys.\ } {\bf 80}, 787–838 (2008).



\bibitem{Festuccia:2011ws}
  G.~Festuccia and N.~Seiberg,
  JHEP {\bf 1106} (2011) 114

\bibitem{deWit:1999ui}
  B.~de Wit, I.~Herger,
 {\it   Lect.\ Notes Phys.\  } {\bf 541}, 79 (2000).



\end{thebibliography}
\end{document}